\documentclass[aoas,nameyear,seceqn,dvips]{arximspdf}
\usepackage{dcolumn}
\usepackage{graphicx}

\doi{10.1214/10-AOAS433}
\volume{5}
\issue{2A}
\pubyear{2011}
\firstpage{924}
\lastpage{942}

\makeatletter
\newtheorem{theorem}{Theorem}[section]
\newcolumntype{d}[1]{D{.}{.}{#1}}
\newcolumntype{k}[1]{D{,}{}{#1}}
\makeatother

\begin{document}
\begin{frontmatter}

\title{Bayesian phase I/II adaptively randomized
oncology trials with combined drugs} %\protect\thanksref{T1}}
\runtitle{Bayesian phase I/II trials with combined drugs}

\begin{aug}
\author[a]{\fnms{Ying} \snm{Yuan}\thanksref{t1}\ead[label=e1]{yyuan@mdanderson.org}}
\and
\author[b]{\fnms{Guosheng} \snm{Yin}\corref{}\thanksref{t2}\ead[label=e2]{gyin@hku.hk}}
\thankstext{t1}{Supported in part by the
National Cancer Institute %(USA)
Grant R01CA154591-01A1.}
\thankstext{t2}{Supported in part by
a grant from the Research Grants Council of Hong Kong.}
\runauthor{Y. Yuan and G. Yin}
\affiliation{University of Texas and University of Hong Kong}

\address[a]{Department of Biostatistics\\
University of Texas\\
MD Anderson Cancer Center\\
Houston, Texas 77030\\
USA \\
\printead{e1}} %\printead*{e2}

\address[b]{Department of Statistics \\
\quad and Actuarial Science\\
University of Hong Kong\\
Hong Kong\\
China\\
\printead{e2}}
\end{aug}

% HISTORY:
\received{\smonth{4} \syear{2010}}
\revised{\smonth{10} \syear{2010}}

% ABSTRACT
%
\begin{abstract}
We propose a new integrated
phase I/II trial design to identify the most efficacious dose combination
that also satisfies certain safety requirements for drug-combination trials.
We first take a Bayesian copula-type model for
dose finding in phase I. After identifying a set of admissible doses,
we immediately move the entire set forward
to phase II. We propose a novel adaptive randomization
scheme to favor assigning patients to more efficacious
dose-combination arms. Our adaptive
randomization scheme takes into account both the point estimate and
variability of efficacy. By using a moving reference to compare the relative
efficacy among treatment arms, our method achieves a high resolution to
distinguish different arms. We also consider groupwise adaptive randomization
when efficacy is late-onset.
We conduct extensive simulation studies to examine the operating
characteristics of the
proposed design, and illustrate our method using a phase I/II melanoma
clinical trial.
\end{abstract}

% KEYWORDS
%

\end{frontmatter}
%
%s1 ###
\section{Introduction}\label{sec1}
Phase I trials usually aim to find the maximum tolerated dose (MTD)
for an investigational drug, and phase II trials examine the efficacy
of the drug at the
identified MTD. Traditionally, phase I and phase
II trials are conducted separately. There is a growing trend to
integrate phase I and phase II trials in order to
expedite the process of drug development and reduce the associated
cost [Gooley et al. (\citeyear{Gooley94}); Thall and Russell (\citeyear
{Thall98});
O'Quigley, Hughes and Fenton (\citeyear{OQuigley01}); Thall and Cook (\citeyear{Thall04});
and Yin, Li and Ji (\citeyear{Yin06}); among others].
The majority of these designs focus on single-agent clinical trials.

Treating patients with a combination of agents is becoming common in cancer
clinical trials. Advantages of such drug-combination
treatments include the potential to induce a synergistic treatment
effect, target tumor cells with differing drug susceptibilities, or
achieve a higher dose intensity with nonoverlapping toxicities.
Trial designs for drug-combination studies involve several
distinct features that are beyond the
scope of methods for single-agent studies. In single-agent
trials, we typically assume that toxicity monotonically increases
with respect to the dose. However, in a drug-combination dose space,
it is difficult to establish such ordering for dose combinations.
Consequently, decision making for dose escalation or
de-escalation is difficult in drug-combination trials due to the
unknown toxicity order.
Another important feature that distinguishes drug-combination trials
from single-agent trials is the toxicity equivalent contour
in the two-dimensional dose-toxicity space. As a result, multiple dose
combinations with similar toxicity
may be found in phase I drug-combination trials.
For these reasons, single-agent phase I/II designs cannot
be directly applied to drug-combination trials.

In spite of a rich body of literature on phase I dose-finding designs for
drug-combination trials [Simon and Korn (\citeyear{Simon90}); Korn and
Simon (\citeyear{Korn93});
Kramar, Lebecq and Candalh (\citeyear{Kramar99}); Thall et al.
(\citeyear{Thall03}); Conaway, Dunbar and Peddada (\citeyear{Conaway04});
Wang and Ivanova (\citeyear{Wang05}); and Yin and Yuan (\citeyear
{Yin09}); among others],
research on phase I/II designs has been very limited. Recently, Huang
et al. (\citeyear{Huang07}) proposed a
parallel phase I/II clinical trial design for combination therapies, which,
however, only targets MTDs with a toxicity probability of 33\%
because the ``3$+$3'' dose-finding
design [Storer (\citeyear{Storer89})] is used in the phase I component.

Our research is motivated by a cancer clinical trial at M. D. Anderson
Cancer Center
for patients diagnosed with malignant melanoma. The experimental agents
to be combined
are decitabine (a DNA methyltransferase inhibitor, which has shown
clinical activity in
patients diagnosed with leukemia or myelodysplastic syndrome)
and a derivative of recombinant interferon
which has been used to treat cancer patients with advanced solid tumors.
The primary objective of the trial is to find the most effective, safe
doses of both drugs
when used in combination to treat melanoma.
For this trial, an integrated phase I/II design is more plausible
to speed up the drug discovery and reduce the total cost.

Toward this goal, we propose a new seamless phase I/II design
to identify the most efficacious dose combination
that also satisfies certain safety requirements for oncology drug-combination
trials. In the phase I part of the trial, we employ a systematic dose-finding
approach by using copula-type regression to model the toxicity of the
drug combinations.
Once phase I is finished, we take a set of admissible doses to phase
II, in which patients are adaptively randomized to multiple treatment arms
corresponding to those admissible doses. We propose a novel adaptive
randomization (AR)
procedure based on a moving reference to compare the relative
efficacy among the treatments in comparison. Our AR has a high
resolution to distinguish
treatments with different levels of efficacy and
thus can efficiently allocate more patients to more efficacious arms.
The proposed design allows us to target any
prespecified toxicity rate and fully utilize the available
information to make dose-assignment decisions.

The rest of the paper is organized as follows. In Section \ref{sec2} we
adopt~a~co\-pula-type probability model for toxicity
and develop a new AR procedure for seamless implementation of the
phase I/II drug-combination trial design. In Section \ref{sec3} we
apply our
design to a melanoma clinical trial, and assess its operating
characteristics through extensive simulation studies. In Section \ref
{sec4} we extend the proposed design to
accommodate trials with late-onset efficacy using group sequential AR.
We conclude with a brief discussion in Section~\ref{sec5}.

%s2 ###
\section{Phase I/II drug-combination design}\label{sec2}
%s2.1 ###
\subsection{Dose finding in phase I} \label{sec2.1}
For ease of exposition,
consider a trial with a combination of two agents, A and B; let $a_i$
be the prespecified toxicity probability corresponding to $A_i$, the
$i$th dose of drug A, with $a_1< a_2 <\cdots<a_I$; and let $b_j$ be that
of $B_j$, the $j$th dose of drug B, with $b_1< b_2 <\cdots<b_J$. Before
the two drugs are combined, each drug should
have been thoroughly investigated when administered alone.
Given the relatively large dose-searching space and the limited sample
size in a drug-combination trial, it is critical to utilize the rich
prior information on $a_i$ and $b_j$ for dose finding.
Typically, the maximum dose for each drug in the
combination is either the individual MTD
determined in the single-agent trials or a dose below the MTD. Therefore,
the specification of $a_i$ and $b_j$ is quite accurate because the upper
bounds $a_I$ and $b_J$ are known.

We employ the copula-type regression in Yin and Yuan (\citeyear{Yin09})
to model
the joint toxicity probability $\pi_{ij}$ at the dose combination
$(A_i, B_j)$,
%e1 ###
%
\begin{equation}
\pi_{ij}=1-\{(1-a_{i}^\alpha)^{-\gamma}+
(1-b_{j}^\beta)^{-\gamma}-1\}^{-1/\gamma}, \label{copulapi}
\end{equation}
where $\alpha, \beta, \gamma> 0$ are unknown model parameters.
This model satisfies the natural constraints for
drug-combination trials. For example, if the toxicity
probabilities of both drugs are zero, the joint toxicity probability is
zero; and if
the toxicity probability of either drug is one, the joint toxicity
probability is one. Another attractive feature of model (\ref{copulapi})
is that if only one drug is tested, it reduces to the well-known continual
reassessment method (CRM) for a single-agent dose-finding design
[O'Quigley, Pepe and Fisher (\citeyear{OQuigley90})].

Although model (\ref{copulapi}) takes a similar functional
form as the Clayton copula [Clayton (\citeyear{Clayton78})],
there are several fundamental differences [Yin and Yuan (\citeyear{Yin10})].
Copula models are widely used to model a bivariate distribution
by expressing the joint probability distribution through the marginal
distributions linked with
a dependence parameter [for example,
see Clayton (\citeyear{Clayton78}); Hougaard (\citeyear{Hougaard86});
Genest and Rivest (\citeyear{Genest93}); and
Nelsen (\citeyear{Nelson06})]. In a drug-combination
trial, we in fact only observe a univariate dose-limiting toxicity
(DLT) outcome for combined agents. For
a patient treated by combined agents $(A_i, B_j)$,
a single binary variable $X$ indicates whether this patient has
experienced DLT:
that is, $X=1$ with probability $\pi_{ij}$, and
$X=0$ with probability $1-\pi_{ij}$. Therefore,
model (\ref{copulapi}) is actually not a copula;
we simply borrow the structure of the Clayton copula to
model the joint toxicity probability when the two drugs are
administered together.
Moreover, model~(\ref{copulapi}) is indexed by three unknown parameters
$(\alpha, \beta, \gamma)$, in which $\gamma$ is similar to the
dependence parameter
in standard copula models and the two extra parameters~$\alpha$ and
$\beta$ render model (\ref{copulapi}) more
flexibility to accommodate the complex two-dimensional
dose-toxicity surface for the purpose of dose finding. Analogous to the
CRM, the parameters $\alpha$ and
$\beta$ also account for the uncertainty of the prespecification of the
single-agent
toxicity probabilities~$a_i$ and $b_j$, thereby enhancing the
robustness of our design to
the misspecification of these prior toxicity probabilities.

Suppose that at a certain stage of the trial, among $n_{ij}$ patients
treated at the paired
doses $(A_i, B_j)$, $x_{ij}$ subjects have experienced DLT. The
likelihood given the observed data $\mathcal{D}$ is
\[
{L}(\alpha, \beta, \gamma| \mathcal{D})\propto
\prod_{i=1}^I\prod_{j=1}^J \pi_{ij}^{x_{ij}}
(1-\pi_{ij})^{n_{ij}-x_{ij}}.
\]
In the Bayesian framework, the joint posterior distribution is given by
\[
f(\alpha, \beta, \gamma|\mathcal{D})\propto
L(\alpha, \beta, \gamma|\mathcal{D})f(\alpha)f(\beta)f(\gamma),
\]
where $f(\alpha)$, $f(\beta)$ and $f(\gamma)$ denote vague gamma prior
distributions
with mean one and large variances for $\alpha$, $\beta$ and $\gamma$,
respectively. We derive the full conditional distributions of these
three parameters and
obtain their posterior samples using the adaptive rejection
Metropolis sampling algorithm [Gilks, Best and Tan (\citeyear{Gilks95})].

%s2.2 ###
\subsection{Adaptive randomization in phase II}\label{sec2.2}
Once the phase I dose finding is complete, the trial seamlessly
moves on to phase II for further efficacy evaluation. Although the main
purpose of phase I
is to identify a set of admissible doses satisfying the safety requirements,
efficacy data are also collected.
Based on the efficacy data collected in both phase I and phase II, each
new cohort of
patients enrolled in phase II are immediately
randomized to a more efficacious treatment arm with a higher probability.
Similar to most of the phase I/II trial designs, patients in phase I
and phase II need
to be homogeneous by meeting certain eligibility criteria, such that
the efficacy data in phase I can be also used to guide adaptive randomization
in phase~II.

For ease of exposition, we
assume that $K$ admissible doses have been found in phase I and will be
subsequently assessed for efficacy using $K$~parallel treatment arms in
phase II.
Let $(p_1, \ldots, p_K)$ denote the response rates corresponding to
the~$K$ admissible doses, and assume that among $n_{k}$ patients
treated in arm $k$, $y_k$ subjects have
experienced efficacy. We model efficacy using the Bayesian
hierarchical model to borrow information across multiple treatment arms:
%e5 ###
%e4 ###
%e3 ###
%e2 ###
%
\begin{eqnarray}\label{Baye}
y_k | p_k &\sim& \mathit{Bi}(n_k, p_k), \nonumber\\
p_k &\sim& \mathit{Be}(\zeta, \xi),
\nonumber
\\[-10pt]
\\[-10pt]
\nonumber
\zeta&\sim& \mathit{Ga}(0.01, 0.01), \\
\xi&\sim& \mathit{Ga}(0.01, 0.01),\nonumber
\end{eqnarray}
where $\mathit{Bi}(n_k, p_k)$ denotes a binomial distribution,
and $\mathit{Be}(\zeta, \xi)$ denotes a~be\-ta distribution with a
shape parameter $\zeta$ and a scale parameter $\xi$. We take vague
gamma prior distributions $\mathit{Ga}(0.01, 0.01)$ with mean one,
for both~$\zeta$ and $\xi$, to ensure that the
data dominate the posterior distribution.
The posterior full conditional distribution of $p_k$ follows
$\mathit{Be}(\zeta+y_k, \xi+n_k-y_k)$, but those of
$\zeta$ and $\xi$ do not have closed forms. As the trial proceeds,
we continuously update the posterior estimates of
the $p_k$'s under model (\ref{Baye}) based on the cumulating data.

The goal of response-AR is to assign
patients to more efficacious treatment arms with higher
probabilities, such that more patients would benefit from
better treatments [Rosenberger and Lachin (\citeyear{Rosenberger02})].
A common practice is to take the assignment probability
proportional to the estimated response rate of each arm, for
example, using the posterior mean of $p_k$ $(k=1, \ldots, K)$.
However, such an AR scheme does not take into
account the variability of the estimated response rates. At the early
stage of a trial, there is only a small amount of data observed, which
would lead to widely spread
and largely overlapping posterior distributions of the $p_k$'s. In this
situation, the estimated response rates of arms 1 and 2, say,
$\hat{p}_1=0.5$ and $\hat{p}_2=0.6$, should not play a dominant role
for patient assignment, because more data are needed to confirm that
arm 2 is truly superior to arm 1. Nevertheless, at a~later stage, after
more patients have been treated and a substantial amount of
data has become available, if we observe $\hat{p}_1=0.5$ and
$\hat{p}_2=0.6$, we would have more confidence in assigning more
patients to arm 2, because its superiority would then be much more strongly
supported. Thus, in addition to the point estimates of the $p_k$'s, their
variance estimates are also critical when determining the randomization
probabilities.

To account for the uncertainty associated with the point
estimates, one can compare the $p_k$'s with a fixed
target, say, $p_0$, and take the assignment probability proportional
to the posterior probability $\operatorname{pr}(p_k>p_0|\mathcal{D})$.
However, in the case where two or more $p_k$'s are much
larger or much smaller than $p_0$, their corresponding posterior
probabilities $\operatorname{pr}(p_k>p_0|\mathcal{D})$ are
either very close to 1 or 0, and, therefore, this AR scheme
would not be able to distinguish them.

Recognizing these limitations of the currently available AR methods,
Huang et al. (\citeyear{Huang07}) arbitrarily took one study treatment
as the reference, say, the first treatment arm, and then
randomized patients based on $R_k=\operatorname{pr}(p_k>p_1|\mathcal
{D})$ for
$k>1$ while setting $R_1=0.5$. For convenience, we refer to this
method as fixed-reference adaptive randomization
(FAR), since each arm is compared with the same fixed reference
to determine the randomization probabilities. By using one of the
treatment arms as the reference, FAR performs better than
that using an arbitrarily chosen target as the reference.
Unfortunately, FAR cannot fully resolve the
problem. For example, in a three-arm
trial if $p_1$ is low but $p_2$ and $p_3$ are high, say,
$p_1=0.1$, $p_2=0.4$ and $p_3=0.6$, FAR may have difficulty
distinguishing arm 2 and arm~3, because both $R_2$ and $R_3$ would be very
close to 1. Even with a sufficient amount of data
to support the finding that arm 3 is the best treatment, the
probabilities of assigning a
patient to arm~2 and arm 3 are still close. This reveals one
limitation of FAR that is due to the use of a fixed reference:
the reference (arm 1) is adequate to distinguish
arm 1 from arms 2 and 3, but may not be helpful to compare arm~2
and arm 3. In addition, because
$R_1=0.5$, no matter how inefficacious arm 1 is, it has an assignment
probability of at least 1$/$5, if we use $R_1/(R_1+R_2+R_3)$ as
the randomization probability to arm 1. Even worse, in the case of a
two-arm trial
with $p_1=0.1$, and $p_2=0.6$, arm 1 has a lower bound of the
assignment probability 1$/$3, which is true even if $p_1=0$ and $p_2=1$.
This illustrates
another limitation of FAR that is due to the direct use of one of
the arms as the reference for comparison. Moreover, the performance of FAR
depends on the chosen reference, with different reference arms leading
to different randomization probabilities.

\begin{figure}

\includegraphics{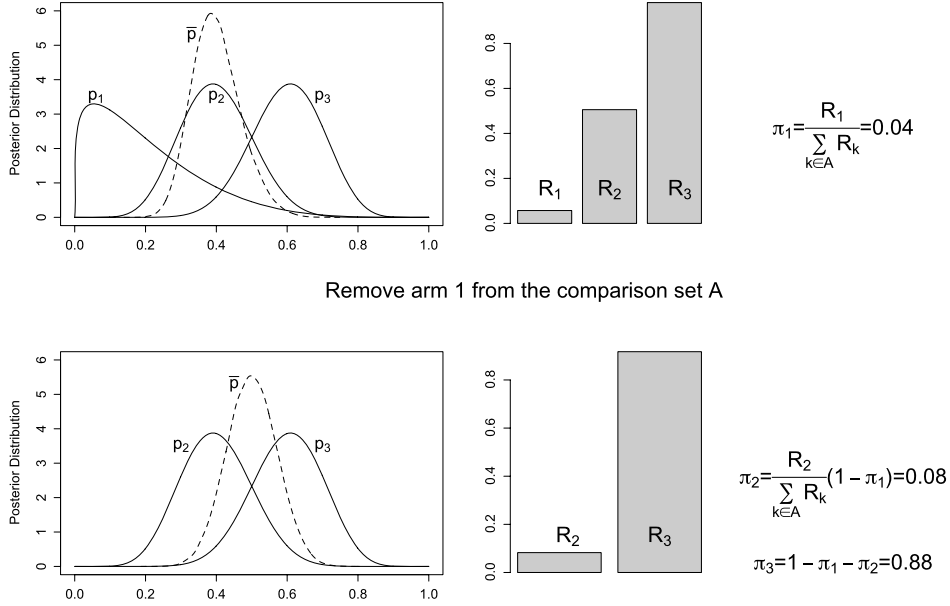}

\caption{Diagram of the proposed moving-reference adaptive randomization
for a three-arm trial. The top panels, from left to right, show that we
first obtain posterior distributions of $p_1, p_2, p_3$
and~$\bar{p}$; then calculate $R_k=\operatorname{pr}(p_k>\bar
{p}|\mathcal{D})$
for $k=1, 2, 3$;
and assign the arm with the smallest value of $R_k$ (i.e., arm 1)
a randomization probability $\pi_1$. After spending $\pi_1$, we remove
arm 1 from
the comparison set and distribute the remaining randomization
probability to the
remaining arms (i.e, arms 1 and 2) in a similar manner, as demonstrated
in the bottom panels. }\label{fig1}
\end{figure}

To fully address the issues with available AR schemes,
we propose a new Bayesian moving-reference adaptive randomization (MAR) method
that accounts for both the magnitude and uncertainty of the estimates
of the $p_k$'s. Unlike FAR, the reference in MAR
is adaptive and varies according to the set of treatment arms under
consideration. One important feature of MAR is
that the set of treatments in comparison is continuously reduced, because
once an arm has been assigned a randomization
probability, it will be removed from the comparison set. By
assigning randomization probabilities to treatment arms
on a one-by-one basis, we can achieve a high resolution to distinguish
different treatments through such a zoomed-in comparison.
Based on the posterior samples of the $p_k$'s,
we diagram the Bayesian MAR in Figure \ref{fig1} and describe it as follows:
\begin{enumerate}
\item Let $\bar\mathcal{A}$ and $\mathcal{A}$ denote the set
of indices of the treatment arms that have and
have not been assigned randomization probabilities, respectively.
We start with $\bar\mathcal{A}=\{\cdot\}$ an empty set,
and $\mathcal{A}=\{1,2, \ldots, K\}$.

\item
Compute the mean response rate for the arms belonging to
the set $\mathcal{A}$, $\bar{p}=\sum_{k \in\mathcal{A}}p_k/\sum_{k \in
\mathcal{A}}1$,
and use $\bar{p}$ as the reference to determine $R_k=\operatorname
{pr}(p_k>\bar
{p}|\mathcal{D})$,
for $k \in\mathcal{A}$. Identify the arm that has the smallest value
of $R_k$, $R_\ell=\min_{k\in\mathcal{A}} R_k$.

\item Assign arm $\ell$ a randomization probability of $\pi_{\ell}$,
\[
\pi_{\ell}=\frac{R_{\ell}}{\sum_{k \in\mathcal{A}} R_{k}}
\biggl(1-\sum_{k' \in\bar\mathcal{A}}\pi_{k'}\biggr),
\]
and update $\mathcal{A}$ and $\bar\mathcal{A}$ by
removing arm $\ell$ from $\mathcal{A}$ into $\bar\mathcal{A}$. Note that
$\pi_{\ell}$ is a fraction of the remaining probability
$1-\sum_{k' \in\bar\mathcal{A}}\pi_{k'}$ because
the assignment probability of $\sum_{k' \in\bar\mathcal{A}}\pi_{k'}$ has
already been ``spent'' in the previous steps.

\item Repeat steps 2 and 3 and keep spending the
rest of the randomization probability until all of the arms are assigned
randomization probabilities, $(\pi_1, \ldots, \pi_K)$, and then randomize
the next cohort of patients to the $k$th arm with a probability of $\pi_k$.
\end{enumerate}
The proposed MAR scheme has a desirable limiting property as given
below.

\begin{theorem}\label{th1}
In a randomized trial with $K$ treatments,
asymptotically, MAR assigns patients to the most efficacious
arm with a limiting probability of 1.
\end{theorem}

The proof is briefly outlined in the \hyperref[app]{Appendix}.
In contrast, using FAR, the probability of
allocating patients to the most efficacious arm may not
converge to 1.

\begin{figure}

\includegraphics{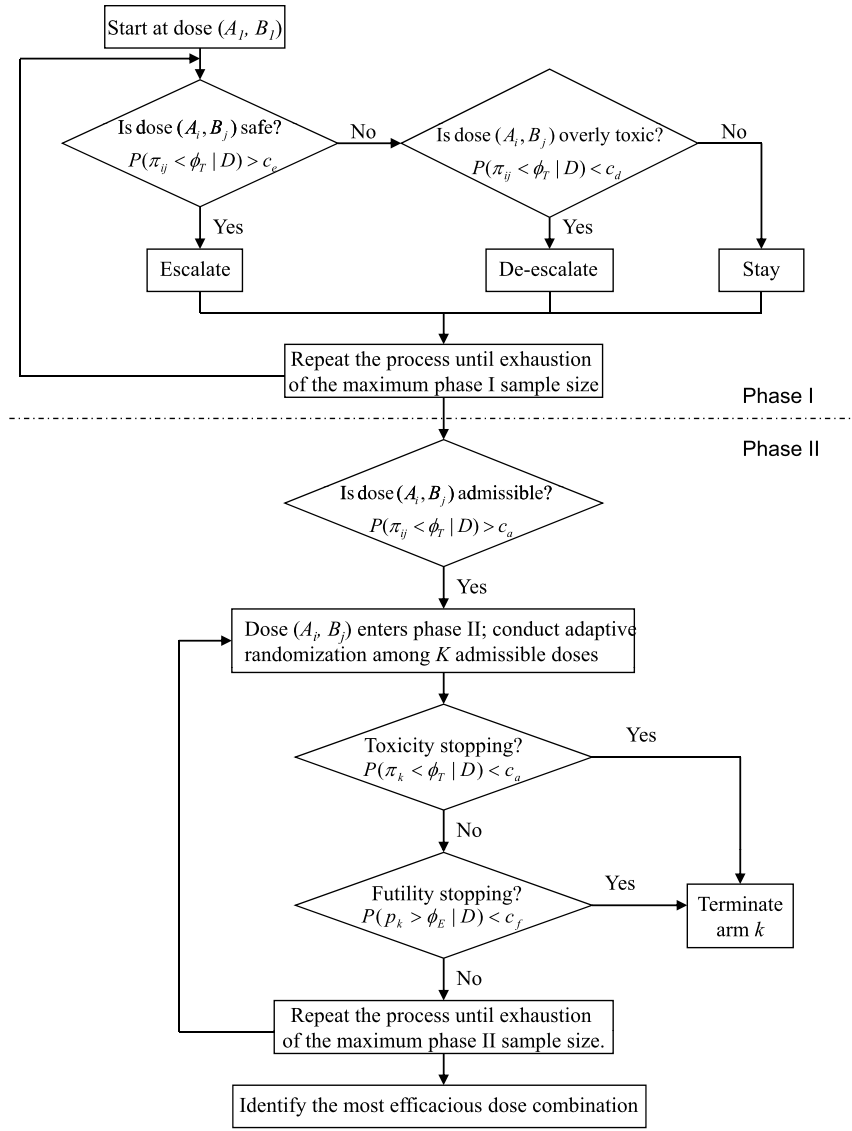}%
\vspace*{-3pt}
\caption{Diagram of the proposed phase I/II trial
design for drug-combination trials.}\label{fig2}
\vspace*{-3pt}
\end{figure}

%s2.3 ###
\subsection{Phase I/II trial design}\label{sec2.3}
The proposed phase I/II
drug-combination design seamlessly integrates each trial
component discussed previously. Let~${\phi_T}$ and $\phi_E$ be the
target toxicity upper limit and efficacy lower limit, and let~$n_1$ and
$n_2$ be the maximum sample sizes for the phase I and phase II parts
of the trial, respectively. Let $c_e$, $c_d$, $c_a$ and $c_f$ be the fixed
probability cutoffs for dose escalation, de-escalation, dose
admissibility and trial futility, the values of which are usually
calibrated through
simulation studies such that the trial has desirable operating
characteristics. Our phase I/II design is displayed in Figure \ref
{fig2} and
described as follows:
\begin{enumerate}
\item In phase I, the first cohort of patients is treated at
the lowest dose combination $(A_1, B_1)$.

\item
During the course of the trial,
at the current dose combination $(A_i, B_j)$:
\begin{enumerate}[(iii)]
\item[(i)]
If $\operatorname{pr}(\pi_{ij}<\phi_T|\mathcal{D})>c_e,$ the doses move
to an
adjacent dose combination chosen from
$\{(A_{i+1}, B_j), (A_{i+1}, B_{j-1}), (A_{i-1}, B_{j+1}),
(A_{i}, B_{j+1})\}$, which has a toxicity probability higher than the
current doses
and closest to $\phi_T$. If the current dose combination is $(A_I,
B_J)$, the
doses stay at the same levels.

\item[(ii)]
If $\operatorname{pr}(\pi_{ij}< \phi_T|\mathcal{D})<c_d,$ the doses move
to an adjacent dose combination chosen from $\{(A_{i-1}, B_j),
(A_{i-1}, B_{j+1}), (A_{i+1}, B_{j-1}),
(A_{i}, B_{j-1})\}$, which has a toxicity probability
lower than the current doses and closest to
$\phi_T$. If the current dose combination is $(A_1, B_1)$, the
trial is terminated.

\item[(iii)]
Otherwise, the next cohort of patients continues to be
treated at the current dose combination.
\end{enumerate}

\item Once the maximum sample size in phase I, $n_1$, is reached,
suppose that there are $K$
dose combinations with toxicity probabilities $\pi_{ij}$ satisfying
$\operatorname{pr}(\pi_{ij}< \phi_T|\mathcal{D})>c_a$, then they
are selected as admissible doses and carried forward to phase II in parallel.

\item In phase II,
MAR is invoked to randomize patients among the $K$ treatment arms.
Meanwhile, the toxicity and futility stopping rules
apply to monitoring each arm: if $\operatorname{pr}(\pi_{k}< \phi
_T|\mathcal{D}) < c_a$ (over-toxic), or $\operatorname{pr}(p_{k}
> \phi_E|\mathcal{D})<c_f$ (futility), arm $k$ is closed, $k=1, \ldots, K$.

\item Once the maximum sample size in phase II, $n_2$,
is reached, the dose combination that has the highest posterior mean
of efficacy is selected as the best dose.
\end{enumerate}

In the proposed design, the response is assumed to be observable
quickly so that each incoming patient can be immediately randomized based
on the efficacy outcomes of previously treated patients. This assumption
can be relaxed by using a group sequential AR approach, when the response
is delayed. The group sequential AR updates the randomization probabilities
after each group of patients' outcomes become available rather
than after each individual outcome [Jennison and Turnbull (\citeyear
{Jennison00})].
Our design is suitable for
trials with a small number of dose combinations, because all the dose
combinations satisfying the safety threshold would be taken into phase II.
If a trial starts with a large number of dose combinations,
many more doses could make it into phase II, possibly some with
toxicity much lower than the upper bound. From a practical point of view,
we could tighten the admissibility criteria by choosing only those
with posterior toxicity probabilities closest to $\phi_T$.

The proposed phase I/II drug-combination trial design has been
implemented using C++. The executable file is available for
free downloading at \href{http://odin.mdacc.tmc.edu/\textasciitilde yyuan/}{http://odin.mdacc.tmc.edu/\textasciitilde yyuan/},
and the
source code is available upon request.\vspace*{2pt}

%s3 ###
\section{Application}\label{sec3}

%s3.1 ###
\subsection{Motivating trial}\label{sec3.1}

We use a melanoma clinical trial to illustrate our phase I/II drug-combination
design. The trial examined
three doses of decita\-bine (drug~A) and two doses of the derivative of
recombinant interferon
(drug~B). The toxicity upper limit was $\phi_T=0.33$, and
the efficacy lower limit was $\phi_E=0.2$. A maximum of 80 patients
were to
be recruited, with $n_1=20$ for phase I, and $n_2=60$ for phase II.
In the copula-type toxicity model, we specified the prior toxicity
probabilities of drug A as $a_i = (0.05, 0.1, 0.2)$,
and those of drug B as $b_j = (0.1, 0.2)$. We elicited %took
%noninformative
prior distributions $\mathit{Ga}(0.5, 0.5)$ for $\alpha$ and $\beta$, and
$\mathit{Ga}(0.1, 0.1)$ for $\gamma$.
The dose-limiting toxicity was
defined as any grade 3 or 4 nonhematologic toxicity, grade 4
thrombocytopenia, or grade 4 neutropenia lasting more than two weeks or
associated with infection. The clinical responses of interest included partial
and complete response.
In this trial, it took up to two weeks to assess both toxicity and efficacy,
rendering the response-adaptive randomization practically
feasible. The accrual rate was two patients per month, and thus
no accrual suspension was needed to wait for
patients' responses in order to assign doses to new patients. It took
approximately
10 months to conduct the phase~I part and
two and a half years to complete the phase II part of the trial.
We used $c_e=0.8$ and $c_d=0.45$ to direct dose escalation and de-escalation,
and $c_a=0.45$ to define the set of admissible doses in phase I.
We applied the toxicity stopping rule of $\operatorname{pr}(\pi_{k}<
\phi_T|\mathcal{D})<c_a$ and the futility stopping rule of
$\operatorname{pr}(p_{k}>
\phi_E|\mathcal{D})<c_f$ with $c_f=0.1$ in phase II. The decisions on
dose assignment and adaptive randomization were made after
observing the outcomes of every individual patient.

%f3 ###
%
\begin{figure}

\includegraphics{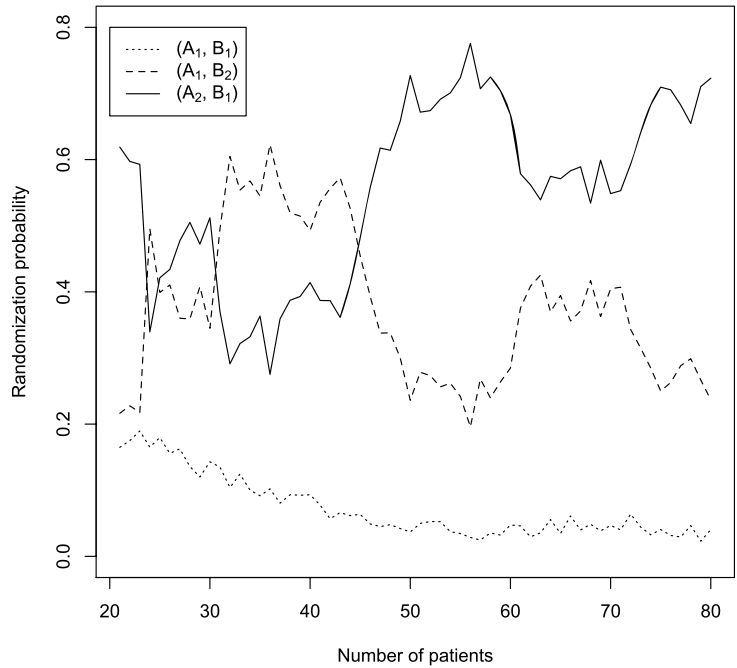}

\caption{Adaptive randomization probabilities for
the three admissible dose combinations in the melanoma clinical
trial.}\label{fig3}
\end{figure}

After 20 patients had been treated in the phase I part of the melanoma trial,
three dose combinations ($A_1, B_1$), ($A_1, B_2$) and
($A_2, B_1$) were identified as admissible doses and carried forward to
phase II for
further evaluation of efficacy. During phase II, the MAR procedure was
used to allocate
the remaining 60 patients to the three dose combinations. Figure \ref
{fig3} displays
the adaptively changing randomization probabilities for
the three treatment arms as the trial proceeded. In particular, the
randomization probability
of ($A_2, B_1$) decreased first, and then increased;
that of ($A_1, B_2$) increased first and
then decreased; and that of ($A_1, B_1$) kept decreasing as the trial
progressed. At the end of the trial, the dose
combination ($A_2, B_1$) was selected as the most desirable dose
with the highest estimated efficacy rate of~0.36.

%s3.2 ###
\subsection{Operating characteristics}\label{sec3.2}
We assessed the operating characteristics of the proposed design via
simulation studies.
Under each of the 12 scenarios given in Table \ref{scenarios}, we simulated 1000 trials.
In the Monte Carlo Markov chain (MCMC) procedure, we
recorded 2000 posterior samples for the model parameters after 100
burn-in iterations.

%t1 ###
%
\begin{table}
\tabcolsep=0pt
\caption{Selection probability and number of patients
treated at each dose combination using the proposed phase I/II design,
with the target dose combinations in boldface}\label{scenarios}
\begin{tabular*}{\textwidth}{@{\extracolsep{\fill
}}lck{1.3}k{1.3}k{1.3}ck{1.3}k{1.3}k{1.3}ck{1.3}k{2.2}k{2.2}ck{2.2}k{2.2}k{2.2}@{}}
\hline
& & \multicolumn{7}{c}{\textbf{Drug A}} && \multicolumn{7}{c}{\textbf{Simulation
results}} \\[-6pt]
& & \multicolumn{7}{c}{\hrulefill} && \multicolumn{7}{c@{}}{\hrulefill}
\\
& & \multicolumn{3}{c}{\textbf{True pr(toxicity)}} && \multicolumn
{3}{c}{\textbf{True pr(efficacy)}} && &&
\\[-6pt]
& \textbf{Drug}&\multicolumn{3}{c}{\hrulefill} && \multicolumn
{3}{c}{\hrulefill} && \multicolumn{3}{c}{\textbf
{Selection}}&&\multicolumn{3}{c}{\textbf{Number of}}\\
\multicolumn{1}{@{}l}{\textbf{Sc.}} & \multicolumn{1}{c}{\textbf{B}}
&\multicolumn{1}{c}{\textbf{1}} & \multicolumn{1}{c}{\textbf{2}} &
\multicolumn{1}{c}{\textbf{3}} & &
\multicolumn{1}{c}{\textbf{1}} & \multicolumn{1}{c}{\textbf{2}} &
\multicolumn{1}{c}{\textbf{3}} & &\multicolumn{3}{c}{\textbf
{percentage}} & &
\multicolumn{3}{c}{\textbf{patients}} \\
\hline
\phantom{0}1&2&0,.1 & 0,.15 & 0,.45 & & 0,.2 & 0,.4 & 0,.6 & & 1,.0 &
25,.2 & 18,.3 & & 8,.8
& 17,.0 & 15,.3 \\
&1&0,.05 & 0,.15 & \textbf{0},\textbf{.2} & & 0,.1 & 0,.3 & \textbf
{0},\textbf{.5} & & 0,.0 &
10,.7 & \textbf{42},\textbf{.8} & & 8,.5 & 11,.3 & \textbf{18},\textbf
{.0} \\
\phantom{0}2&2&0,.1 & \textbf{0},\textbf{.2} & 0,.5 & & 0,.2 & \textbf
{0},\textbf{.4} & 0,.55 & & 4,.0 &
\textbf{44},\textbf{.5} & 2,.8 & & 11,.3 & \textbf{21},\textbf{.2} &
8,.1\\
&1& 0,.05 & 0,.15 & 0,.4 & & 0,.1 & 0,.3 & 0,.5 & & 0,.3 & 24,.0 &
19,.2 & & 9,.7
& 15,.8 & 11,.4 \\
\phantom{0}3&2& 0,.1 & 0,.15 & \textbf{0},\textbf{.2} & & 0,.2 & 0,.3 &
\textbf{0},\textbf{.5} & & 1,.7 &
7,.0 & \textbf{67},\textbf{.1} & & 8,.3 & 10,.9 & \textbf{31},\textbf
{.3} \\
&1& 0,.05 & 0,.1 & 0,.15 & & 0,.1 & 0,.2 & 0,.4 && 0,.0 & 1,.9 & 19,.8
& & 8,.2 &
7,.9 & 11,.9\\
\phantom{0}4&2&0,.1 & 0,.4 & 0,.6 & & 0,.3 & 0,.5 & 0,.6 & & 16,.3 &
25,.4 & 0,.2 & & 16,.1
& 15,.1 & 3,.7 \\
&1&0,.05 & \textbf{0},\textbf{.2} & 0,.5 & & 0,.2 & \textbf{0},\textbf
{.4} & 0,.55 & & 3,.9 &
\textbf{46},\textbf{.2} & 3,.1 && 14,.2 & \textbf{22},\textbf{.3} &
5,.7 \\
\phantom{0}5&2& 0,.1 & \textbf{0},\textbf{.2} & 0,.25 & & 0,.3 & \textbf
{0},\textbf{.5} & 0,.2 && 7,.6 &
\textbf{52},\textbf{.8} & 0,.3 & & 11,.1 & \textbf{19},\textbf{.9} &
13,.4 \\
&1& 0,.05 & 0,.15 & 0,.2 & & 0,.2 & 0,.4 & 0,.4 && 0,.8 & 20,.5 & 15,.4
& & 10,.2
& 12,.5 & 11,.2 \\
\phantom{0}6&2&0,.05 & 0,.05 & \textbf{0},\textbf{.05} & & 0,.2 & 0,.3
& \textbf{0},\textbf{.5} && 0,.3 &
5,.2 & \textbf{77},\textbf{.8} & & 7,.5 & 9,.6 & \textbf{37},\textbf
{.9} \\
&1&0,.05 & 0,.05 & 0,.05 & & 0,.1 & 0,.2 & 0,.4 & & 0,.0 & 0,.3 & 16,.0
& & 7,.7
& 6,.7 & 10,.4\\
\phantom{0}7&2 & 0,.1 & \textbf{0},\textbf{.2} & 0,.5 & & 0,.2 & \textbf
{0},\textbf{.4} & 0,.5 && 4,.2 &
\textbf{41},\textbf{.8} & 9,.3 & & 10,.6 & \textbf{20},\textbf{.3} &
10,.7\\
&1& 0,.05 & 0,.15 & \textbf{0},\textbf{.2} & & 0,.1 & 0,.3 & \textbf
{0},\textbf{.4} && 0,.5 &
10,.7 & \textbf{29},\textbf{.8} & & 9,.5 & 12,.1 & \textbf{15},\textbf
{.0} \\
\phantom{0}8&2&0,.5 & 0,.55 & 0,.6 & & 0,.5 & 0,.5 & 0,.5 & & 0,.0 &
0,.0 & 0,.0 & &0,.4 &
0,.3 & 0,.1 \\
&1&0,.5 & 0,.55 & 0,.6 & & 0,.5 & 0,.5 & 0,.5 & & 0,.1 & 0,.0 & 0,.0 &
& 7,.3 &
0,.4 & 0,.1 \\
\phantom{0}9&2& 0,.4 & 0,.72 & 0,.9 & & 0,.44 & 0,.58 & 0,.71 & & 0,.5
& 0,.0 & 0,.0 & &
3,.6 & 1,.6 & 0,.3 \\
&1& \textbf{0},\textbf{.23} & 0,.4 & 0,.59 && \textbf{0},\textbf{.36} &
0,.49 & 0,.62 & &
\textbf{23},\textbf{.9} & 3,.7 & 0,.0 & & \textbf{20},\textbf{.9} &
6,.0 & 0,.8 \\
10&2&0,.24 & 0,.56 & 0,.83 && 0,.4 & 0,.6 & 0,.78& & 10,.8 & 2,.5 &
0,.0 & &11,.6
& 6,.0 & 1,.2\\
&1&0,.13 & \textbf{0},\textbf{.25} & 0,.42 && 0,.32 & \textbf{0},\textbf
{.5} & 0,.68& & 19,.0 &
\textbf{41},\textbf{.6} & 3,.3 & &22,.1 & \textbf{19},\textbf{.9}&
4,.0\\
11 &2& 0,.15 & \textbf{0},\textbf{.25} & 0,.4 && 0,.3 & \textbf
{0},\textbf{.41} & 0,.54 & &17,.1
& \textbf{35},\textbf{.4} & 17,.2 & & 13,.3 & \textbf{16},\textbf{.5} &
12,.8 \\
& 1& 0,.11 & 0,.15 & 0,.2 && 0,.15 & 0,.22 & 0,.31 & & 1,.6 & 7,.3 &
11,.3 & &
10,.8 & 10,.3 & 9,.8 \\
12&2&0,.15 & 0,.19 & \textbf{0},\textbf{.23} && 0,.17 & 0,.33 & \textbf
{0},\textbf{.55} & & 1,.6 &
10,.1 & \textbf{54},\textbf{.1} & & 7,.7 & 11,.3 & \textbf{25},\textbf
{.6} \\
&1&0,.12 & 0,.15 & 0,.19 && 0,.1 & 0,.22 & 0,.39& & 0,.3 & 4,.2 & 17,.9
& &9,.0 &
8,.2 & 10,.5\\
\hline
\end{tabular*}
\end{table}

In each scenario, the target dose-combination is defined as
the most efficacious one belonging to the admissible set.
We present the selection percentages and the numbers of patients
treated at all of the dose combinations.
Scenarios 1--4 represent the most common cases in which both
toxicity and efficacy increase with the dose levels, while
the target doses are located differently in the two-dimensional space.
The target dose is $(A_3, B_1)$ in scenario~1, and $(A_2, B_2)$ in
scenario 2, which not only had the
highest selection probability, but was also the arm to which most of
the patients were randomized. In scenario 3
the target dose is the combination of the highest doses of
drug A and drug B, for which the selection probability
was close to 70\%, and more than 30 patients
were treated at the most efficacious dose.
Scenario 4 also demonstrated a good performance of
our design with a high selection probability of the target dose.
In scenario 5 toxicity increases with the dose but efficacy
first increases then decreases, and in scenario 6
toxicity maintains at a very low level, but efficacy gradually
increases with the dose.
Under these two scenarios,
both the selection probabilities and the numbers of patients allocated
to the target dose
were plausible. Scenario 7 has two target doses due to the toxicity
and efficacy equivalence contours. In that scenario, both the target
doses were selected
with much higher percentages and more patients were assigned to those
two doses than others.
Scenario 8 demonstrated the safety of our design by
successfully terminating the trial early when
toxicity is excessive even at the lowest dose.
Scenarios 9--12 are constructed for a sensitivity analysis, which
will be described in Section \ref{sec3.3}.

%t2 ###
%
\begin{table}
\tabcolsep=0pt
\caption{Selection percentage of each dose combination to the
admissible set and the average size of the admissible set, $\bar
{K}$, in phase I.
The true admissible doses are in boldface}\label{tab2}
\begin{tabular*}{\textwidth}{@{\extracolsep{\fill}}lcccc ccccc cccc@{}}
\hline
\multicolumn{3}{c}{\textbf{\% of admissible}} & $\bolds{\bar{K}}$ &&
\multicolumn{3}{c}{\textbf{\% of admissible}} & $\bolds{\bar{K}}$ &&
\multicolumn{3}{c}{\textbf{\% of admissible}} & $\bolds{\bar{K}}$ \\
\hline
\multicolumn{3}{c}{Scenario 1} &&& \multicolumn{3}{c}{Scenario 2} &&&
\multicolumn{3}{c}{Scenario 3} \\
\textbf{97.8} &\textbf{95.2} & 67.9 & 5.4 &&
\textbf{97.9} &\textbf{92.2} & 38.2 & 4.9 &&
\textbf{99.6} & \textbf{99.0} &\textbf{91.8} & 5.9 \\
\textbf{98.1} & \textbf{98.0} &\textbf{91.9} & &&
\textbf{98.8} &\textbf{98.2} & 72.6 &&&
\textbf{99.7} & \textbf{99.7} &\textbf{98.3} &\\
\multicolumn{3}{c}{Scenario 4} &&& \multicolumn{3}{c}{Scenario 5} &&&
\multicolumn{3}{c}{Scenario 6} \\
\textbf{93.3} & 75.9 & 14.6 & 4.2 && \textbf{98.0} & \textbf{95.9}
&\textbf{81.6} & 5.6 &&
\textbf{99.5} & \textbf{99.5} &\textbf{99.0} & 6.0 \\
\textbf{96.5} & \textbf{94.9} & 52.1 & &&
\textbf{98.6} & \textbf{98.5} &\textbf{93.7} & &&
\textbf{99.5} & \textbf{99.5} &\textbf{99.5} & \\
\multicolumn{3}{c}{Scenario 7} &&& \multicolumn{3}{c}{Scenario 8} &&&
\multicolumn{3}{c}{Scenario 9} \\
\textbf{97.9} &\textbf{94.6} & 55.6 & 5.3 &&
\phantom{0}2.0 & \phantom{0}0.9 & \phantom{0}0.3 & 0.1 && 24.9 & 10.5 &
\phantom{0}0.4 & 1.2 \\
\textbf{98.8} & \textbf{98.5} &\textbf{86.1} & && \phantom{0}3.1
& \phantom{0}2.4 & \phantom{0}0.9 & && \textbf{42.7} & 36.6 & 11.6 & \\
\multicolumn{3}{c}{Scenario 10} &&& \multicolumn{3}{c}{Scenario 11}
&&& \multicolumn{3}{c}{Scenario 12} \\
\textbf{69.6} & 42.7 & \phantom{0}4.9 & 3.1 && \textbf{88.4} &\textbf
{82.5} & 53.5 & 4.8
&& \textbf{91.9} & \textbf{89.3} &\textbf{73.4} & 5.3 \\
\textbf{82.4} &\textbf{78.3} & 36.4 & &&
\textbf{91.1} & \textbf{90.1} &\textbf{77.7} & &&
\textbf{92.5} & \textbf{92.4} &\textbf{87.6} & \\
\hline
\end{tabular*}
\end{table}

To better understand the performance of the phase I part of
the proposed design, in Table \ref{tab2} we display the percentage
of each
dose being selected into the admissible set,
and the average number of admissible doses, $\bar K$, at the end of
phase I.
In most of the cases,
the selection percentages of the admissible doses were higher than 90\%, and
the average number of admissible doses determined by the proposed design
was close to the true value. For example, in scenario 1, the
true number of admissible doses is 5, and our design, on average,
selected 5.4 admissible doses for further study in phase II.

%t3 ###
%
\begin{table}
\tabcolsep=0pt
\caption{Number of patients randomized to each treatment arm using the
fixed-reference adaptive randomization (FAR) compared to the
moving-reference adaptive randomization (MAR). The most efficacious
dose is in boldface}\label{tab3}
\begin{tabular*}{\textwidth}{@{\extracolsep{\fill
}}lk{1.3}k{1.3}ccck{2.2}k{2.2}ck{2.2}k{2.2}k{2.2}@{}}
\hline
& \multicolumn{3}{c}{\textbf{Response rate}}
& &\multicolumn{3}{c}{\textbf{FAR}} &&\multicolumn{3}{c}{\textbf{MAR}}
\\[-6pt]
& \multicolumn{3}{c}{\hrulefill}
& &\multicolumn{3}{c}{\hrulefill} &&\multicolumn{3}{c@{}}{\hrulefill} \\
\textbf{Sc.} &\multicolumn{1}{c}{\textbf{Arm 1}} & \multicolumn
{1}{c}{\textbf{Arm 2}} &\multicolumn{1}{c}{\textbf{Arm 3}} &&
\multicolumn{1}{c}{\textbf{Arm 1}} & \multicolumn{1}{c}{\textbf{Arm 2}}
&\multicolumn{1}{c}{\textbf{Arm 3}} & & \multicolumn{1}{c}{\textbf{Arm
1}} & \multicolumn{1}{c}{\textbf{Arm 2}}
&\multicolumn{1}{c@{}}{\textbf{Arm 3}} \\
\hline
1 & 0,.1 & 0,.2 & \textbf{0.3} & & 27.7 & 31,.8 & \textbf{40},\textbf
{.6} & & 12,.5 &
29,.0 & \textbf{58},\textbf{.5} \\
2 & 0,.2 & 0,.1 & \textbf{0.3} & & 40.5 & 17,.2 & \textbf{42},\textbf
{.4} & & 27,.3 &
13,.0 & \textbf{59},\textbf{.7} \\
3 & \textbf{0},\textbf{.3} & 0,.1 & 0.2 & & \textbf{61.1} & 13,.7 &
25,.2 & & \textbf{58},\textbf{.4} & 13,.1 & 28,.5 \\
4 & 0,.1 & 0,.3 & \textbf{0.6} & & 23.3 & 33,.8 & \textbf{42},\textbf
{.9} & & 5,.5 &
13,.3 & \textbf{81},\textbf{.3} \\
5 & 0,.3 & \textbf{0},\textbf{.6} & 0.1 & & 34.4 & \textbf{54},\textbf
{.6} & 11,.0 & & 13,.9 &
\textbf{80},\textbf{.5} & 5,.5 \\
6 & \textbf{0},\textbf{.6} & 0,.3 & 0.1 & &\textbf{82.2} & 12,.5 & 5,.3
& & \textbf{81},\textbf{.8} & 12,.8 & 5,.3 \\
7 & 0,.01 & 0,.4 & \textbf{0.6} & & 21.0 & 38,.7 & \textbf{40},\textbf
{.3} & &
3,.7 & 20,.6 & \textbf{75},\textbf{.7} \\
8 & 0,.01 & 0,.01 & \textbf{0.5} & & 25.8 & 25,.1 & \textbf{49},\textbf
{.1} & & 5,.3
& 5,.3 & \textbf{89},\textbf{.4} \\
\hline
\end{tabular*}
\end{table}

\begin{figure}

\includegraphics{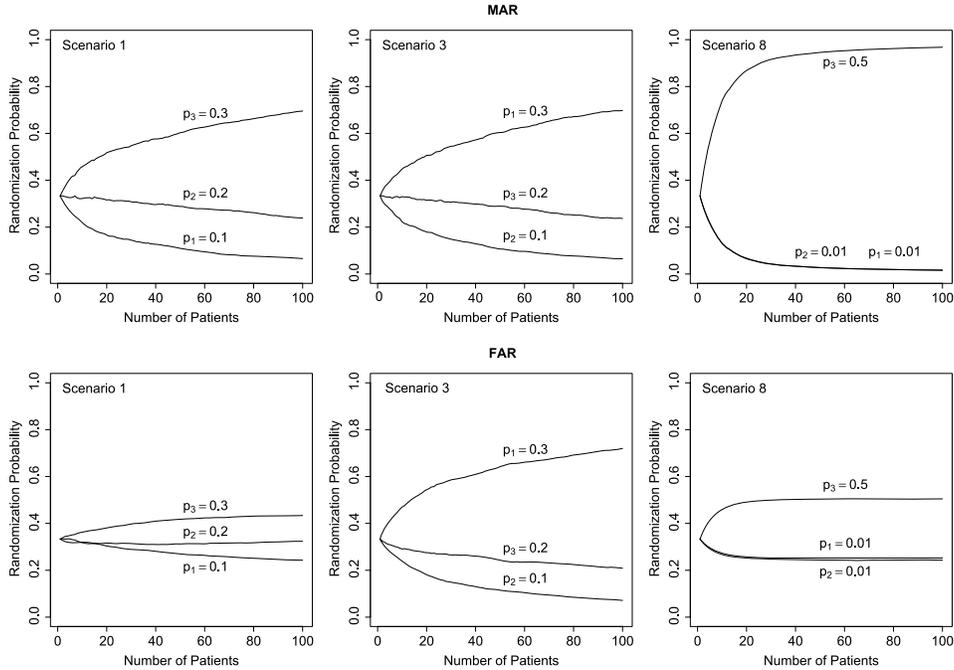}%
\vspace*{-3pt}
\caption{Randomization probabilities of
the proposed moving-reference adaptive randomization (MAR)
and the fixed-reference adaptive randomization
(FAR) under scenarios 1, 3 and 8 listed in Table \protect\ref{tab3}.}\label{fig4}
\vspace*{-3pt}
\end{figure}

As the number of admissible doses selected by phase I
may vary from one trial to another, the final trial
results shown in Table \ref{scenarios} are jointly affected by both the
phase I and
phase II
parts of the design. To disassemble their intertwining
effects, we conducted a simulation study with a focus on the adaptive
randomization only.
In particular, we considered a phase II trial in which a total
of 100 patients would be randomized to three treatment arms.
Table \ref{tab3} shows the results based on 1000 simulated trials under
eight different scenarios. Scenarios 1--3 simulate cases in which the first
arm has the lowest, intermediate and highest efficacy,
respectively. Scenarios 4--6 are constructed in a similar setting,
but the efficacy differences
among the three arms are much larger. Scenarios 7 and 8 consider
cases in which one or two arms are futile. In all of the scenarios,
MAR allocated the majority of
patients to the most efficacious arm in a more efficient way
than FAR. For~scena\-rios~1, 3 and 8, in Figure \ref{fig4} we show
the randomization probabilities averaged over 1000 simulations with
respect to the
cumulative number of patients using MAR and FAR, respectively.
As more data are collected,
MAR has a~substantially higher resolution to distinguish and separate
treatment arms than FAR in terms of efficacy. For example, in scenario 1
the curves are adequately separated using MAR
after 20 patients are randomized, but are still
not well spread even after enrolling 40 patients using FAR.
Furthermore, considering scenarios 4, 5 and~6, we see that the number of
patients assigned to the most efficacious arm (with a response
rate of 0.6) using FAR changed substantially from 42.9 to 82.2,
whereas that number stayed approximately the same as 81 when using MAR.
This phenomenon indicates the invariance of MAR
and the sensitivity of FAR to the reference arm.

%
%f4 ###
%

%s3.3 ###
\subsection{Sensitivity analysis}\label{sec3.3}
In the first sensitivity analysis,
we examined the robustness of the proposed design to model
misspecifications. We generated true toxicity and
efficacy probabilities from the logistic regression model,
%e6 ###
%
\begin{equation}\label{logit}
\pi_{ij}=\frac{\exp(\beta_0+\beta_1Z_{Ai}+\beta_2
Z_{Bj}+\beta_3Z_{Ai}Z_{Bj})}{1+\exp(\beta_0+\beta_1Z_{Ai}+\beta_2
Z_{Bj}+\beta_3Z_{Ai}Z_{Bj})},
\end{equation}
but applied models (\ref{copulapi}) and (\ref{Baye}) for estimation.
We took the standardized doses of drugs A and B in model
(\ref{logit}) as $Z_{Ai}=(0.05, 0.1, 0.2)$ and $Z_{Bj}=(0.1, 0.2)$.
These cases are listed as scenarios 9--12 in Table \ref{scenarios}.
When the models for toxicity and efficacy were misspecified,
our design still performed very well: the target dose combination
was selected with the highest probability and most of the patients were
allocated to those efficacious dose combinations.

In the second sensitivity analysis, we evaluated the impact of the
prior specifications
using two more diffusive prior distributions for $\alpha$ and
$\beta$ under scenarios 1--4. The simulation results in Table \ref{tab4}
using the more diffusive priors are very close
to those for scenarios 1--4 in Table \ref{scenarios}.
Therefore, the proposed design does not appear to be sensitive to
the prior specification.

%s4 ###
\section{Late-onset efficacy}\label{sec4}
In practice, toxicity and efficacy
outcomes need to be ascertainable shortly after the initiation of the
treatment in order to make a real-time decision on the
treatment assignment for each incoming patient.
Often, toxicity can be observed quickly; whereas
efficacy is late-onset, for example, tumor shrinkage may take a
relatively long time to assess.
Such delayed efficacy outcomes pose new challenges to the use of AR in
randomized trials. We propose using the group sequential AR procedure,
which adapts randomization probabilities after a group
of patients' outcomes become available rather than after observing each
individual's outcome [Karrison, Huo and Chappell (\citeyear
{Karrison03})]. More specifically, for
the $n_2$ patients to be randomized to $K$ treatment arms in phase II,
we update the AR probabilities after observing every $m$ patients'
outcomes, $1\le m\le n_2$. Choosing an appropriate $m$ is critical
for the practical performance of the group sequential AR.
With a larger value of $m$, the trial duration tends to be shortened
because we suspend the accrual less frequently, but it may
downgrade the AR performance.
Using a smaller group size $m$, the group sequential AR procedure would
better facilitate assigning more patients to
more efficacious treatment arms, but it prolongs the trial duration.
In addition to the group size,
the performance of the design also depends on the accrual rate,
the length of the follow-up required for efficacy assessment
and the distribution of the time to efficacy.

%t4 ###
%
\begin{table}
\caption{Sensitivity analysis of the proposed Bayesian phase I/II
drug-combination design under different prior specifications}\label{tab4}
\tabcolsep=0pt\vspace*{-3pt}
\fontsize{7.8}{9.8}\selectfont{
\begin{tabular*}{\textwidth}{@{\extracolsep{4in minus 4in}}lcccccccccccc@{}}
\hline
& \multicolumn{6}{c}{\textbf{Selection percentage}} & \multicolumn{6}{c@{}}{\textbf{Number
of treated patients}} \\[-6pt]
%& \multicolumn{6}{c}{\rule{167pt}{0.5pt}} && \multicolumn{6}{c@{}}{\rule{154pt}{0.5pt}} \\
& \multicolumn{6}{c}{\hrulefill} & \multicolumn{6}{c@{}}{\hrulefill} \\
\textbf{Sc.} & \multicolumn{3}{c}{$\bolds{\alpha, \beta\sim{Ga(0.1, 0.1)}}$} &
\multicolumn{3}{c}{$\bolds{\alpha, \beta\sim{Ga(0.05, 0.05)}}$} &
\multicolumn{3}{c}{$\bolds{\alpha, \beta\sim{Ga(0.1, 0.1)}}$} & \multicolumn
{3}{c@{}}{$\bolds{\alpha, \beta\sim
{Ga(0.05, 0.05)}}$} \\
\hline
1 & \phantom{0}2.0 & 21.0 & 25.4  & \phantom{0}2.0 & 19.9 & 27.1  & \phantom{0}7.8 & 15.7 & 20.2  &
\phantom{0}7.7 & 15.7 & 20.8\\
& \phantom{0}0.1 & \phantom{0}8.2 & \textbf{41.8} & 0 & \phantom{0}9.0 & \textbf{40.2}  & \phantom{0}7.4 & \phantom{0}9.6
& \textbf{18.3} & \phantom{0}7.1 & \phantom{0}9.6 & \textbf{18.1} \\
2 & \phantom{0}5.4 & \textbf{46.5} & \phantom{0}2.9  & \phantom{0}4.9 & \textbf{45.9} & \phantom{0}3.7  & 10.6 &
\textbf{21.6} & \phantom{0}9.6  & 10.5 & \textbf{21.4} & \phantom{0}9.4\\
& \phantom{0}0.4 & 21.0 & 20.3 & \phantom{0}0.2 & 22.9 & 17.9 & \phantom{0}8.8 & 14.4 & 13.0 &
\phantom{0}8.8 & 14.9 & 12.5 \\
3 & \phantom{0}2.0 & \phantom{0}6.3 & \textbf{69.6} & \phantom{0}0.9 & \phantom{0}6.1 & \textbf{70.2}  & \phantom{0}7.2 &
10.1 & \textbf{36.3} & \phantom{0}6.8 & \phantom{0}9.8 & \textbf{36.4}\\
& \phantom{0}0.1 & \phantom{0}1.2 & 19.4 & 0 & \phantom{0}1.4 & 19.9  & \phantom{0}7.1 & \phantom{0}6.5 & 12.1 & \phantom{0}7.0 &
\phantom{0}6.6 & 12.5 \\
4 & 16.9 & 24.7 & \phantom{0}0.2 & 15.6 & 24.6 & \phantom{0}0.3  & 15.2 & 16.4 & \phantom{0}4.0 &
15.1 & 16.0 & \phantom{0}4.3\\
& \phantom{0}4.8 & \textbf{44.2} & \phantom{0}2.8 & \phantom{0}4.1 & \textbf{45.1} & \phantom{0}3.0  & 13.1 &
\textbf{20.6} & \phantom{0}6.8 & 13.2 & \textbf{20.4} & \phantom{0}6.2 \\
\hline
\end{tabular*}}
\vspace*{-6pt}
\end{table}

To evaluate our design using the group sequential AR,
we took efficacy to be late-onset, requiring three months for a
complete evaluation.
We considered six different group sizes:
$m={}$1, 3, 6, 12, 20 and 30, corresponding to 1.7\%, 5\%, 10\%, 20\%,
33.3\%
and 50\% of the total sample size, $n_2=60$, in phase II.
We investigated two different accrual rates: two and eight patients per month,
and simulated four different patterns of the hazard for the time to efficacy:
increasing, constant, decreasing and hump-shaped over time.
The first three hazards were generated from the Weibull
distribution, and the hump-shaped hazard was generated from the
log-logistic distribution. Other design parameters, such as $\phi_T$,
$c_e$, $c_d$ and $c_a$,
took the same values as those in Section \ref{sec3.1}.

%t5 ###
%
\begin{table}
\tabcolsep=0pt
\caption{Number of patients allocated to the target
dose combination and the trial duration (shown as subscripts),
with different group sizes $m$ under scenarios 1--5}\label{tab5}
\vspace*{-5pt}
\begin{tabular*}{\textwidth}{@{\extracolsep{4in minus 4in}}l lllllllll@{}}
\hline
& \multicolumn{4}{c}{\textbf{Hazard (accrual rate${} \bolds{=}
{}$2/month)}} && \multicolumn
{4}{c@{}}{\textbf{Hazard (accrual rate${} \bolds{=} {}$8/month)}} \\[-6pt]
& \multicolumn{4}{c}{\hrulefill} && \multicolumn
{4}{c@{}}{\hrulefill} \\
$\bolds{m}$ & \multicolumn{1}{c}{\textbf{Increase}} & \multicolumn
{1}{c}{\textbf{Constant}} & \multicolumn{1}{c}{\textbf{Decrease}} &
\multicolumn{1}{c}{\textbf{Hump}} & &
\multicolumn{1}{c}{\textbf{Increase}} & \multicolumn{1}{c}{\textbf
{Constant}} &
\multicolumn{1}{c}{\textbf{Decrease}} & \multicolumn{1}{c@{}}{\textbf
{Hump}} \\
\hline
\multicolumn{10}{@{}l}{Scenario 1} \\
\phantom{0}1 & 18.4$_{ 157.8}$ & 18.1$_{ 150.3}$ & 17.7$_{ 139.5}$ & 17.8$_{
159.9}$ && 18.2$_{ 149.4}$ & 17.5$_{ 143.0}$ & 17.8$_{ 131.5}$ &
17.4$_{ 152.2}$ \\
\phantom{0}3 & 18.0$_{ 69.8}$ & 18.0$_{ 69.2}$ & 17.5$_{ 68.7}$ & 17.9$_{ 69.4}$
&& 17.2$_{ 61.6}$ & 17.7$_{ 60.9}$ & 17.9$_{ 60.7}$ & 17.5$_{ 61.5}$ \\
\phantom{0}6 & 17.7$_{ 42.4}$ & 17.4$_{ 42.4}$ & 16.9$_{ 42.4}$ & 17.3$_{ 42.2}$
&& 17.6$_{ 33.0}$ & 17.1$_{ 32.8}$ & 17.3$_{ 32.8}$ & 16.7$_{ 32.9}$ \\
12 & 16.5$_{ 42.2}$ & 16.8$_{ 42.2}$ & 16.1$_{ 42.1}$ & 16.5$_{ 42.0}$
&& 16.6$_{ 18.8}$ & 16.6$_{ 18.7}$ & 16.9$_{ 18.8}$ & 16.9$_{ 18.7}$ \\
20 & 15.8$_{ 42.1}$ & 16.0$_{ 42.1}$ & 15.8$_{ 42.1}$ & 15.7$_{ 42.1}$
&& 15.6$_{ 13.8}$ & 16.3$_{ 13.8}$ & 16.1$_{ 13.8}$ & 15.4$_{ 13.7}$ \\
30 & 15.0$_{ 42.1}$ & 15.4$_{ 42.1}$ & 15.0$_{ 42.0}$ & 15.2$_{ 42.0}$
&& 15.0$_{ 12.8}$ & 15.1$_{ 12.8}$ & 15.3$_{ 12.8}$ & 14.3$_{
12.8}$\\[3pt]
\multicolumn{10}{@{}l}{Scenario 2} \\
\phantom{0}1 & 21.0$_{ 160.2}$ &20.7$_{ 153.7}$ & 22.5$_{ 145.3}$ & 21.9$_{
162.3}$ && 21.3$_{ 151.9}$ & 21.7$_{ 146.9}$ & 22.0$_{ 135.8}$ &
21.2$_{ 154.5}$ \\
\phantom{0}3 & 20.8$_{ 69.5}$ & 21.6$_{ 69.4}$ & 21.9$_{ 68.6}$ & 21.5$_{ 69.2}$
&& 21.3$_{ 61.3}$ & 21.2$_{ 61.2}$ & 21.5$_{ 60.8}$ & 20.9$_{ 61.1}$ \\
\phantom{0}6 & 20.7$_{ 42.3}$ & 21.1$_{ 42.1}$ & 21.7$_{ 42.2}$ & 20.9$_{ 42.1}$
&& 20.7$_{ 32.8}$ & 21.6$_{ 32.7}$ & 21.8$_{ 32.6}$ & 20.9$_{ 32.6}$ \\
12 & 20.5$_{ 42.0}$ & 20.8$_{ 42.0}$ & 21.2$_{ 42.1}$ & 20.9$_{ 41.9}$
&& 20.5$_{ 18.6}$ & 20.8$_{ 18.6}$ & 21.0$_{ 18.7}$ & 20.6$_{ 18.7}$ \\
20 & 20.2$_{ 42.0}$ & 19.1$_{ 42.1}$ & 20.2$_{ 42.0}$ & 20.8$_{ 42.1}$
&& 20.0$_{ 13.7}$ & 19.3$_{ 13.7}$ & 20.1$_{ 13.8}$ & 19.6$_{ 13.7}$ \\
30 & 19.1$_{ 42.2}$ & 19.6$_{ 42.1}$ & 19.6$_{ 41.9}$ & 19.7$_{ 42.1}$
&& 18.6$_{ 12.8}$ & 19.6$_{ 12.8}$ & 18.5$_{ 12.8}$ & 19.3$_{ 12.8}$
\\[3pt]
\multicolumn{10}{@{}l}{Scenario 3} \\
\phantom{0}1 & 32.2$_{ 161.2}$ &32.0$_{ 154.6}$ & 31.7$_{ 144.7}$ & 31.1$_{
163.6}$ && 31.9$_{ 153.8}$ & 32.0$_{ 147.1}$ & 32.1$_{ 136.6}$ &
32.3$_{ 156.5}$ \\
\phantom{0}3 & 31.5$_{ 70.1}$ & 31.8$_{ 69.7}$ & 31.4$_{ 69.2}$ & 31.5$_{ 69.9}$
&& 31.0$_{ 61.8}$ & 31.3$_{ 61.7}$ & 32.0$_{ 61.2}$ & 31.5$_{ 61.9}$ \\
\phantom{0}6 & 32.4$_{ 42.4}$ & 31.0$_{ 42.4}$ & 32.2$_{ 42.3}$ & 30.9$_{ 42.4}$
&& 31.2$_{ 33.0}$ & 31.0$_{ 33.0}$ & 32.4$_{ 32.9}$ & 31.5$_{ 32.9}$ \\
12 & 30.8$_{ 42.2}$ & 31.0$_{ 42.2}$ & 30.3$_{ 42.1}$ & 30.3$_{ 42.3}$
&& 30.4$_{ 18.8}$ & 30.6$_{ 18.8}$ & 31.4$_{ 18.8}$ & 31.9$_{ 18.8}$ \\
20 & 29.5$_{ 42.3}$ & 30.4$_{ 42.1}$ & 29.6$_{ 42.2}$ & 30.1$_{ 42.1}$
&& 28.7$_{ 13.8}$ & 28.2$_{ 13.8}$ & 28.2$_{ 13.8}$ & 29.0$_{ 13.8}$ \\
30 & 29.4$_{ 42.2}$ & 29.1$_{ 42.2}$ & 28.9$_{ 42.2}$ & 29.1$_{ 42.2}$
&& 26.4$_{ 12.8}$ & 26.7$_{ 12.8}$ & 26.4$_{ 12.8}$ & 26.5$_{ 12.8}$
\\[3pt]
\multicolumn{10}{@{}l}{Scenario 4} \\
\phantom{0}1 & 23.0$_{ 158.4}$ & 23.7$_{ 151.5}$ & 22.8$_{ 139.4}$ & 22.4$_{
159.5}$ && 23.5$_{ 150.4}$ & 22.7$_{ 143.8}$ & 22.3$_{ 131.3}$ &
22.6$_{ 151.9}$ \\
\phantom{0}3 & 22.7$_{ 69.1}$ & 22.3$_{ 68.9}$ & 22.6$_{ 68.7}$ & 22.4$_{ 69.1}$
&& 22.3$_{ 61.0}$ & 22.8$_{ 60.9}$ & 21.9$_{ 59.8}$ & 21.0$_{ 61.2}$ \\
\phantom{0}6 & 21.8$_{ 42.4}$ & 21.7$_{ 42.2}$ & 22.5$_{ 42.3}$ & 21.3$_{ 42.3}$
&& 22.2$_{ 32.6}$ & 21.9$_{ 32.8}$ & 22.4$_{ 32.9}$ & 21.5$_{ 32.7}$ \\
12 & 21.2$_{ 42.1}$ & 21.3$_{ 41.9}$ & 20.7$_{ 42.1}$ & 21.0$_{ 41.8}$
&& 20.4$_{ 18.6}$ & 22.0$_{ 18.7}$ & 21.9$_{ 18.7}$ & 21.2$_{ 18.7}$ \\
20 & 20.2$_{ 42.1}$ & 20.3$_{ 42.1}$ & 19.8$_{ 42.0}$ & 20.1$_{ 42.0}$
&& 20.0$_{ 13.7}$ & 20.1$_{ 13.7}$ & 19.9$_{ 13.7}$ & 20.3$_{ 13.8}$ \\
30 & 19.3$_{ 42.2}$ & 19.3$_{ 42.1}$ & 18.6$_{ 42.1}$ & 19.4$_{ 42.2}$
&& 19.5$_{ 12.8}$ & 19.2$_{ 12.7}$ & 20.2$_{ 12.8}$ & 18.9$_{ 12.8}$
\\[3pt]
\multicolumn{10}{@{}l}{Scenario 5} \\
\phantom{0}1 & 20.4$_{ 159.7}$ & 21.3$_{ 152.1}$ & 19.8$_{ 141.1}$ & 20.3$_{
161.6}$ && 20.5$_{ 152.4}$ & 20.5$_{ 144.7}$ & 21.1$_{ 133.4}$ &
20.9$_{ 154.1}$ \\
\phantom{0}3 & 20.0$_{ 70.1}$ & 20.6$_{ 69.9}$ & 20.4$_{ 69.1}$ & 20.1$_{ 69.9}$
&& 20.4$_{ 61.8}$ & 20.0$_{ 61.5}$ & 20.3$_{ 61.1}$ & 19.6$_{ 61.9}$ \\
\phantom{0}6 & 19.9$_{ 42.5}$ & 19.6$_{ 42.5}$ & 19.9$_{ 42.5}$ & 20.0$_{ 42.5}$
&& 19.5$_{ 33.0}$ & 19.6$_{ 32.9}$ & 20.0$_{ 33.0}$ & 20.5$_{ 33.0}$ \\
12 & 19.2$_{ 42.2}$ & 19.3$_{ 42.3}$ & 19.7$_{ 42.2}$ & 20.0$_{ 42.3}$
&& 19.1$_{ 18.8}$ & 19.2$_{ 18.8}$ & 18.7$_{ 18.8}$ & 19.6$_{ 18.8}$ \\
20 & 19.5$_{ 42.3}$ & 19.5$_{ 42.3}$ & 18.8$_{ 42.2}$ & 18.8$_{ 42.3}$
&& 18.3$_{ 13.8}$ & 18.8$_{ 13.8}$ & 18.6$_{ 13.8}$ & 18.8$_{ 13.8}$ \\
30 & 18.5$_{ 42.3}$ & 18.1$_{ 42.2}$ & 19.1$_{ 42.2}$ & 18.4$_{ 42.3}$
&& 17.2$_{ 12.8}$ & 17.6$_{ 12.8}$ & 17.7$_{ 12.8}$ & 17.7$_{ 12.8}$ \\
\hline
\end{tabular*}
\vspace*{-6pt}
\end{table}

Table \ref{tab5} shows the number of patients allocated to the target
dose combination
and the duration of the trial under the first five scenarios listed in
Table \ref{scenarios}.
In general, when the size of the sequential group $m$
increases, the number of patients allocated to the target dose combination
gradually decreases. This phenomenon was minor when $m$
increased from 1 to 6, but more notable when $m$ became larger.
For example, in scenario 1 with an increasing hazard and
an accrual rate of two patients per month,
the numbers of patients allocated to the target dose combination
were 18.4, 17.7 and 15.0, when $m=1, 6$ and 30, respectively.
The trial duration was more sensitive to the value of $m$,
and changed dramatically when the size of the sequential group increased.
When the accrual rate was two
patients per month, we observed a substantial decrease in the trial
duration when $m$ increased from 1 to~6. For example,
under scenario 2, the duration of the trial with
$m=6$ was approximately 1$/$4 of that with $m=1$. However, when we
further increased~%
$m$ from~6 to 30, the trial duration only changed slightly because
in this circumstance the trial duration was essentially dominated
by the accrual rate, which is typically a key factor affecting the
trial duration. With a higher accrual rate of eight patients
per month, we observed additional reductions in the trial duration when
$m$ was
larger than 6, but as a trade-off, slightly fewer
patients were allocated to the target dose combination.

In practice, trial duration is an important factor to be considered when
designing clinical trials. We should choose an appropriate
group size so as to achieve a reasonable balance between
AR and the trial duration. It is worth noting that the group size is
mainly used to determine when
to update the randomization probabilities, not when to randomize patients.
Patients are randomized on a one-by-one basis
to the treatment arms no matter the size of the sequential group. In
the extreme case that the group size equals the total
sample size, we essentially apply an equal randomization scheme.

%s5 ###
\section{Concluding remarks}\label{sec5}
Drug-combination therapies are playing an increasingly important role
in oncology
research. Due to the toxicity equivalence contour in the
two-dimensional dose-combination space, multiple dose combinations
with similar toxicity may be identified in a phase I trial. Thus,
a phase II trial with AR is natural and ethical to assign more
patients to more efficacious doses.
We have adopted a copula-type model to select the admissible
doses and proposed a novel AR scheme when
seamlessly connecting phase I and phase II trials.
The attractive feature of this phase I/II design is that
once the admissible doses are identified,
AR immediately takes effect based on the efficacy data collected in the
phase I study. The proposed design efficiently uses all of the
available data
resources and naturally bridges the phase I and phase II trials.
In our design, AR is based only on efficacy comparison among admissible doses.
It can be easily modified to take into account
both toxicity and efficacy by using their odds ratio
as a measure of the trade-off or desirability
to adaptively randomize patients and select the best dose combination
at the end of the trial [Yin, Li and Ji (\citeyear{Yin06})].
The proposed design assumes that both toxicity and efficacy
endpoints are binary. In some cases,
these endpoints can be ordinal or continuous, for example,
it may be more direct to treat the toxicity endpoint
as an ordinal outcome to account for multiple toxicity grades.
To accommodate such an ordinal toxicity outcome, we can take the
approach of Yuan, Chappell and Bailey (\citeyear{Yuan07})
by first converting the toxicity grades to numeric scores that reflect
their impact
on the dose allocation procedure, and then incorporating
those scores into the copula-type model using the quasi-binomial likelihood.
Another common scenario is that both toxicity and efficacy endpoints take
the form of time-to-event measurements.
In this case, various survival models, such as the proportional hazards
model, are available to model the times to toxicity and efficacy.
Along this direction, Yuan and Yin (\citeyear{Yuan09}) discussed
jointly modeling
toxicity and efficacy as time-to-event outcomes in single-agent trials;
similar approaches can be adopted here for phase I/II drug-combination
trials.

\begin{appendix}

%s6 ###
\section*{\texorpdfstring{Appendix: Proof of Theorem \lowercase{\protect\ref{th1}}}
{Appendix: Proof of Theorem 2.1}}\label{app}

The proposed moving-reference adaptive randomization procedure is
invariant to the labeling of the treatment arms.
Without loss of generality, we assume that $p_1<p_2<\cdots<p_K$, and
determine the randomization probability for arm 1. Starting
with $\mathcal{A}=\{1, 2, \ldots, K\}$,
$\bar{p}=\sum_{k=1}^K p_k/K$.
As the number of subjects goes to infinity, it follows that the rank of
$R_k= \operatorname{pr}(p_k>\bar{p}|\mathcal{D}) $ is consistent with
the order of the $p_k$'s, that is, $R_1<R_2<\cdots<R_K$. Therefore,
for the first treatment arm $\ell=1$,
\begin{eqnarray*}
R_{\ell} &=& \min_{k\in\mathcal{A}} R_k \\
&=& \operatorname{pr}(p_1>\bar{p}|\mathcal{D}) \\
&=& \operatorname{pr}\{ (p_1-p_{2})+(p_1-p_{3})+\cdots+(p_1-p_{K})
>0|\mathcal
{D}\} \\
& \le& \operatorname{pr}\{ (K-1)(p_1-p_{2}) >0|\mathcal{D}\}\\
& = & \operatorname{pr}( p_1>p_{2} |\mathcal{D}),
\end{eqnarray*}
which converges to 0 asymptotically.
Thus, $\pi_1 \to0$, that is, the probability of assigning patients
to the least efficacious arm goes to zero. Following similar arguments,
we can show that
$\pi_k \to0$ for $k=2, \ldots, K-1$. Therefore, the probability of
allocating patients
to the most efficacious arm,
$\pi_K=1-\sum_{k=1}^{K-1}\pi_k$, converges to 1.
\end{appendix}

\section*{Acknowledgments}
We would like to
thank the referees, Associate Editor and Editor (Professor Karen
Kafadar) for very helpful comments
that substantially improved this paper.

%suskaldyti doi

\printaddresses

\end{document}